\documentclass[11pt]{article}
\usepackage{graphicx}
\usepackage{amsmath,amsthm,amssymb,amsfonts,mathtools}
\begin{document}

\begin{titlepage}
{\ }
\vskip 1in

\begin{center}
{\LARGE{Generalized symmetries and 2-groups \\ \vskip0.05in via electromagnetic duality in AdS/CFT}}
\vskip 0.5in {\Large Oliver DeWolfe\footnote{oliver.dewolfe@colorado.edu} and Kenneth Higginbotham\footnote{kenneth.higginbotham@colorado.edu}}
\vskip 0.2in {\it Department of Physics and \\
Center for Theory of Quantum Matter \\
390 UCB \\ University of Colorado \\ Boulder, CO 80309, USA}
\end{center}
\vskip 0.5in

\begin{abstract}\noindent
We discuss how electromagnetically dualizing a 1-form to a 2-form in AdS$_5$ exchanges regular and alternate boundary conditions, and thus gauges the originally global $U(1)$ symmetry in the dual field theory. The generalized symmetry current dual to the 2-form in the bulk is identified as the dual field strength of the gauged $U(1)$, and the associated double-trace operator with a logarithmically running coupling is just the gauged $U(1)$ Maxwell action. Applying this dualization to an AdS Maxwell-Chern-Simons theory dual to a global $U(1) \times U(1)$ model with an 't Hooft anomaly results in a theory with a modified field strength that holographically realizes a 2-group symmetry. We explicitly carry out the holographic renormalization to verify this, and discuss the generalization to other rank fields in other dimensions.
\end{abstract}

\end{titlepage}

\section{Introduction}

Symmetries are a fundamental aspect of physics, and their association with conserved charges and currents is a deep principle. In recent years, it has become increasingly apparent that it is useful to think about non-scalar conserved quantities beyond the familiar energy-momentum. These higher-rank conserved quantities are associated with so-called generalized symmetries, which have become the subject of much study since the work of Gaiotto, Kapustin, Seiberg, and Willett \cite{Gaiotto:2014kfa}. An elementary example occurs with a $U(1)$ gauge field in four dimensions, where the dual field strength behaves as a two-index generalized current that is conserved in the absence of magnetic sources due to the Bianchi identity. 

Novel symmetry structures may also arise when these generalized symmetries interplay with ordinary symmetries. C\'ordova, Dumitrescu and Intriligator have described how gauging certain global symmetries with 't Hooft anomalies leads to a generalized symmetry that for consistency must also transform nontrivially under ordinary symmetries of the theory, producing a so-called 2-group structure \cite{Cordova:2018cvg}. This 2-group structure can illuminate aspects of the field theory, and is generalizable to cases with higher rank symmetries. Earlier developments of 2-group symmetries include \cite{baez2003higher, Baez:2004in, Baez:2005sn, Baez:2005, Sati:2008eg, Schreiber2008,  Schreiber2008-2, Kapustin:2013uxa}, while more recent work on 2-groups includes \cite{Tachikawa:2017gyf, Benini:2018reh, Cordova:2020tij,  DelZotto:2020sop, Hidaka:2020iaz, Hidaka:2020izy, Yu:2020twi}. 

Meanwhile, the AdS/CFT correspondence is a powerful way to understand quantum field theories outside of the weak coupling limit by means of a ``holographic" dual gravitational description living in asymptotically anti-de Sitter space \cite{Maldacena:1997re, Gubser:1998bc, Witten:1998qj}. The quantum field theory ``lives" at the boundary of the bulk spacetime, and the boundary limits of bulk fields are associated to field theory operators and their sources, while holographic renormalization to remove divergences is carried out by adding counterterms living at the boundary. Symmetry properties of the field theory have natural realizations in the dual gravity picture; spacetime symmetries are mapped to isometries of the metric, and global symmetries are mapped to symmetries of fluctuating gauge fields. It is natural that higher-rank generalized symmetries are realized in the gravity dual by dynamical $p$-form gauge fields.

The holographic dual of a 2-form current in four dimensional conformal field theories is thus a 2-form gauge field in AdS$_5$, and this was studied by Grozdanov and Poovuttikul \cite{Grozdanov:2017kyl} and Hofman and Iqbal \cite{Hofman:2017vwr}, who noticed that the solutions to the ordinary Maxwell action for the 2-form involve logarithmic behavior, which is not compatible with conformal invariance \cite{Grozdanov:2017kyl, Hofman:2017vwr}. This logarithmic behavior is associated with the possibility of adding to the field theory action a classically marginal ``multi-trace" term bilinear in the 2-form current, whose coupling runs logarithmically. More work on generalized symmetries and AdS/CFT includes \cite{Grozdanov:2018ewh, Harlow:2018tng, Bergman:2020ifi}.

In this paper we explore this issue further, and find an interpretation of the logarithmic behavior of the 2-form current. In the AdS/CFT correspondence, the boundary conditions on bulk fields determine the identity of the dual field theory operator. One set of ``regular" boundary conditions is always possible, but for certain fields one may have additional ``alternate" boundary conditions where what was the source now becomes the dual operator, leading to an entirely different dual quantum field theory. An ordinary 1-form gauge field in the bulk with regular boundary conditions is dual to the current of a conserved global $U(1)$ symmetry, with a non-dynamical background gauge field as its source. However, were the 1-form gauge field given alternate boundary conditions this background gauge field would become dynamical, and the dual theory would describe a gauged $U(1)$. This was noted in the context of AdS$_4$ by Witten \cite{Witten:2003ya}, where the phenomenon was related to bulk electromagnetic duality.

We observe that in five dimensions, a 1-form gauge field $A_\mu$ and a 2-form $B_{\mu\nu}$ are electromagnetic duals of each other, and thus represent the same bulk physics. We may ask how this relates to the physics in the dual field theory. We indeed find that the electromagnetic duality exchanges regular and alternate boundary conditions. Thus the 2-form gauge field with regular boundary conditions describes the dual physics of a 1-form with alternate boundary conditions, that is, a dynamical $U(1)$ gauge field. Indeed, the electromagnetic duality map reveals that the conserved two-index generalized symmetry current in the field theory is nothing but the dual field strength of this gauged $U(1)$. Moreover, the associated multi-trace operator is shown to be simply the Maxwell action for this gauged $U(1)$, and thus the running coupling that breaks conformal symmetry is the ordinary coupling of a QED-like theory coupled to electric matter. Related observations, with an application to magnetohydrodynamics, are made in \cite{Grozdanov:2017kyl}.

After going through the case of the 1-form and the 2-form in AdS$_5$ in some detail, we show how the statement that electromagnetic duality in AdS exchanges regular and alternate boundary conditions, and thus gauges the dual global symmetry, holds for general rank $p$-forms in general dimension. This is the gravity dual of the statement that gauging any conserved generalized symmetry produces a new generalized symmetry from the Bianchi identity of the new dynamical gauge field.

We also examine an application of this relationship. In \cite{Cordova:2018cvg}, the expected gravity dual for a theory with a 2-group symmetry was identified as a theory of a 2-form and a 1-form where the 2-form has a modified field strength. Also in \cite{Cordova:2018cvg} a simple example of a field theory with a 2-group structure was given, by starting with a $U(1) \times U(1)$ global symmetry  with an 't Hooft anomaly, and gauging one of the $U(1)$s; the new generalized symmetry from the gauge field combines with the remaining global symmetry to produce a 2-group. In AdS/CFT, the global symmetry model is realized by a pair of gauge fields in AdS$_5$ with a Chern-Simons coupling capturing the 't Hooft anomaly. We show that performing electromagnetic duality to replace one of the gauge fields with a 2-form, thereby gauging the associated $U(1)$, leads precisely to the gravity theory with modified field strength expected to capture the 2-group structure, and verify in detail using holographic renormalization that the 2-group structure is indeed present. We also discuss the natural generalization to $n$-group structure in arbitrary dimension. Other work on AdS/CFT and 2-groups with a focus on hydrodynamics recently appeared in \cite{Iqbal:2020lrt}.

In section~\ref{OneFormTwoFormSec} we recapitulate the ideas of boundary conditions and holographic renormalization in AdS/CFT, and study the 1-form and 2-form in AdS$_5$ in detail, showing how their electromagnetic duality relation exchanges boundary conditions  and gauges the global symmetry, before discussing generalizations to other forms in other dimensions at end of the section. In section~\ref{ChernSimonsModfiedSec}, we review the $U(1) \times U(1)$ 2-group model of \cite{Cordova:2018cvg}, and show how electromagnetic duality on the gravity dual Chern-Simons theory produces the holographic dual to a theory with the 2-group symmetry, before commenting on the generalization to $n$-groups. Two flowcharts are provided to summarize the relations discussed. In section~\ref{ConclusionsSec} we conclude.

\section{Electromagnetic duality, boundary conditions and gauging global symmetries}
\label{OneFormTwoFormSec}

In anti-de Sitter space, one must specify boundary conditions for a dynamical field at spatial infinity, and moreover the choice of boundary conditions affects the identity of the corresponding operator in the dual field theory and its dynamics. Rendering the action finite and compatible with the boundary conditions involves the addition of appropriate boundary counterterms, a process called holographic renormalization.

We first review this for the more familiar case of a scalar field, before turning to our primary subjects of study, the 1-form $A_\mu$ and 2-form $B_{\mu\nu}$ in AdS$_5$, and their relationship via electromagnetic duality. We will see how  alternate boundary conditions for the gauge field $A_\mu$, which describe a dynamical $U(1)$ field in the dual field theory, can be realized by ordinary boundary conditions for the magnetic dual field $B_{\mu\nu}$. This 2-form field is dual to a 2-form conserved current, which is shown to be just the magnetic field strength of the dynamical $U(1)$ gauge field. Double-trace boundary conditions are possible given an appropriate choice of finite counterterm, which we identity as the Maxwell action for this dynamical $U(1)$.

\subsection{Review of holographic renormalization}

Consider a field $\phi(\vec{x},t,r)$ living in $(d+1)$-dimensional anti-de Sitter space, whose metric in the Poincar\'e patch we take to be
\begin{eqnarray}
	ds^2 = g_{\mu\nu} dx^\mu dx^\nu = {r^2 \over L^2} \eta_{ij} dx^i dx^j + {L^2 \over r^2} dr^2 \,,
\end{eqnarray}
where $L$ is the AdS length scale, Greek indices denote $(d+1)$-dimensional coordinates in the bulk, while Roman indices denote $d$-dimensional coordinates on the boundary (which exclude $r$), and $\eta_{ij}$ is the $d$-dimensional Minkowski metric. Solutions to the Klein-Gordon equation ${1 \over\sqrt{-g}} \partial_\mu \sqrt{-g} g^{\mu\nu} \partial_\nu \phi = m^2 \phi $ near the boundary look like
\begin{eqnarray}
\label{PhiSoln}
	\phi(\vec{x},t,r \to \infty) = \alpha(\vec{x},t){ L^{2\Delta_-} \over r^{\Delta_-}} + \ldots + \beta(\vec{x},t){L^{2\Delta_+} \over r^{\Delta_+}} + \ldots \,, \qquad \Delta_\pm \equiv {d \over 2} \pm \sqrt{\left(d \over 2\right)^2 + m^2 L^2} \,,
\end{eqnarray}
where $\alpha(\vec{x},t)$ and $\beta(\vec{x},t)$ are the leading terms in each independent solution. Using the scaling isometry $\vec{x} \to \zeta\vec{x}, t \to \zeta t, r \to r/\zeta$ we can see that $\alpha(\vec{x}, t)$ has scaling dimension $\Delta_-$ and $\beta(\vec{x}, t)$ has scaling dimension $\Delta_+$; we have inserted factors of $L$ in (\ref{PhiSoln}) to make the engineering dimensions match the scaling dimensions.

The bulk action evaluated on solutions to the equations of motion generally diverges. This can be remedied by cutting off the bulk space at some large radial coordinate $r=R$, with induced boundary metric
\begin{eqnarray}
	ds^2 = h_{ij} dx^i dx^j = {R^2 \over L^2} \eta_{ij} dx^i dx^j \,,
\end{eqnarray}
and adding boundary terms to the action that cancel the divergences as the cutoff is removed $R \to \infty$. In general, the finite parts of these boundary counterterms can be adjusted freely. This process is the bulk realization of the process of regularizing and renormalizing a quantum field theory, and goes by the name ``holographic renormalization" \cite{Bianchi:2001de, Bianchi:2001kw, deHaro:2000vlm, Skenderis:2002wp} 
(for additional review, see \cite{DeWolfe:2018dkl, Papadimitriou:2004ap}; for additional developments, see \cite{Papadimitriou:2004ap, Papadimitriou:2010as}).

Moreover, boundary conditions for fields in AdS are naturally associated to the choice of terms in the boundary action. We demand the action be stationary on solutions to the equations of motion, and while the bulk action will always be so, in general the boundary part of the action is only stationary when suitable boundary conditions are chosen. For a scalar $\phi(\vec{x},t,r)$ of generic mass $m^2$, the natural choice is $\alpha(\vec{x},t)$ fixed, the so-called regular boundary condition. In this case $\beta(\vec{x},t)$ is identified as the expectation value of the dimension-$\Delta_+$ operator ${\cal O}_{\rm reg}$ dual to $\phi$,
\begin{eqnarray}
	\langle {\cal O}_{\rm reg}(\vec{x},t)\rangle \sim \beta(\vec{x},t) \,,
\end{eqnarray}
 while $\alpha(\vec{x},t)$ is its source. A boundary counterterm of the form $S_{\rm ct} \sim \int d^dx \sqrt{-h}\, \phi^2$ renders the action finite and stationary with these regular boundary conditions.

However, for scalars fields with mass in the range $- d^2/4 < m^2 L^2 \leq - d^2/4 + 1 $, it is possible to make a different choice, the alternate boundary condition \cite{Klebanov:1999tb}. Adding a different set of boundary terms, the action can be made finite and stationary for the boundary condition $\beta(\vec{x},t)$ fixed, and then $\beta(\vec{x},t)$ acts as a source to an operator ${\cal O}_{\rm alt}(\vec{x},t)$ with dimension $\Delta_-$, whose expectation value is
\begin{eqnarray}
	\langle {\cal O}_{\rm alt}(\vec{x},t)\rangle \sim \alpha(\vec{x},t) \,.
\end{eqnarray}
We see that a change in boundary conditions (and correspondingly to the boundary action) changes the operator content of the dual quantum field theory.

More general boundary conditions are also possible, as a condition like $\alpha(\vec{x},t) \propto \beta(\vec{x},t)$ is associated to the insertion into the field theory action of the ``double-trace" operator $\Delta S_{\rm CFT} \propto \int d^dx \, {\cal O}_{\rm reg}(\vec{x},t)\,{\cal O}_{\rm reg}(\vec{x},t)$. In general the addition of the multi-trace operator $W[{\cal O}_{\rm reg}]$ to the field theory gives a boundary condition of the form $\alpha(\vec{x},t) \propto \delta W[\beta] / \delta \beta(\vec{x},t)$ \cite{Aharony:2001pa,  Witten:2001ua,Berkooz:2002ug}, see also \cite{Papadimitriou:2007sj}.

\subsection{The one-form $A_\mu$ in AdS$_5$ and its dual global $U(1)$ symmetry}

Consider now the Maxwell action for a 1-form field $A_\mu$ in five dimensions ($d=4$):
\begin{eqnarray}
\label{AAction}
	S_{\rm A} = \int d^5x \sqrt{-g} \left( - {1 \over 4} F_{\mu\nu} F^{\mu\nu} \right) \,,
\end{eqnarray}
with $F_{\mu\nu} = \partial_\mu A_\nu - \partial_\nu A_\mu$.
The bulk equation of motion is
\begin{eqnarray}
	\partial_\mu \sqrt{-g} F^{\mu \nu} =0 \,,
\end{eqnarray}
which becomes
\begin{eqnarray}
\eta^{jk} \partial_j F_{k r} =0 \,, \quad \quad
\left(	\partial_r + {3 \over r} \right) F_{r i} + {L^4\over r^4} \eta^{jk} \partial_j F_{ki} = 0 \,.
\label{pFormEOM2}
\end{eqnarray}
In the gauge $A_r = 0$, the solution to leading order is
\begin{eqnarray}
\label{ASoln}
	A_i(\vec{x},t, r) = \alpha_i(\vec{x},t) L + \gamma_i(\vec{x},t){ L^5 \log r \over r^2} + \beta_i(\vec{x},t){ L^5 \over r^2} + \ldots \,.
\end{eqnarray}
The functions $\alpha_i(\vec{x},t)$ and $\beta_i(\vec{x},t)$ are the two independent modes, with scaling dimensions 1 and 3, respectively.
$\beta_i(\vec{x},t)$ is constrained by the first equation of motion to satisfy\footnote{In our notation naturally 4-dimensional quantities like $\alpha_i$ and $\beta_i$ have their indices raised and lowered by the Minkowski metric $\eta_{ij}$.}
\begin{eqnarray}
\label{BetaDivergence}
	\partial_i \beta^i = 0 \,,
\end{eqnarray}
 while the equations of motion determine $\gamma_i(\vec{x},t)$ completely as
\begin{eqnarray}
	\gamma^i = {1 \over 2} \partial_j f^{ji} \,,
\end{eqnarray}
with $f_{ij} \equiv \partial_i \alpha_j - \partial_j \alpha_i$. Residual gauge transformations of $A_\mu$ are $r$-independent, and become transformations of $\alpha_i(\vec{x},t)$:
\begin{eqnarray}
\label{AGaugeInvariance}
	\alpha_i(\vec{x},t) \to \alpha_i(\vec{x},t) + \partial_i \lambda(\vec{x},t) \,.
\end{eqnarray}
Plugging in the solution (\ref{ASoln}) to the action, the bulk action vanishes, but we are left with a boundary term. Cutting off AdS at some large $r = R$ we find
\begin{eqnarray}
	S_{\rm A} &=& - {L^3 \over 2} \int d^4x \,\alpha_i \left[ (1 - 2 \log R)\gamma^i - 2 \beta^i \right] \,, \\
	&=&  L^3 \int d^4x \, \left[ \alpha_i \beta^i + \left(\log R - {1 \over 2}\right)\left(- {1 \over 4} f_{ij}f^{ij} \right) \right]  \,, \nonumber
\end{eqnarray}
where we have dropped terms vanishing in the $R \to \infty$ limit, and in the second line we integrated by parts to cast the latter term in the form of a Maxwell action for $\alpha_i$. Furthermore, when we vary the field by $\delta A_i = \delta \alpha_i L + \delta \gamma_i \log r/r^2 + \delta \beta_i /r^2 + \ldots$, we obtain the variation of the action as a boundary term,
\begin{eqnarray}
	\delta S_{\rm A} &=& - L^3  \int d^4x\, \delta \alpha_i \left[ (1 - 2 \log R)\gamma^i - 2 \beta^i \right]\\
	&=&  2L^3  \int d^4x \, \left[ \delta\alpha_i \beta^i + \left(\log R - {1 \over 2}\right)\left(- {1 \over 4} \delta f_{ij}f^{ij} \right) \right]  \,. \nonumber
\end{eqnarray}
We want both the action and its variation to be finite on the equations of motion, but so far they are divergent due to the $\log R$ factor. To address this we add a boundary counterterm,
\begin{eqnarray}
\label{ACounterterm}
	S_{{\rm ct}, A} = L \int d^4x \sqrt{-h} \left( - {1 \over 4 e^2(R)} F_{\mu\nu} F^{\mu\nu} \right)\,,
\end{eqnarray}
where we noted how the coefficient $1/e^2(R)$ may depend on the value of the cutoff $R$. For the choice
\begin{eqnarray}
\label{ECoefficient}
	{1 \over e^2(R)} =  {1 \over 2} - \log R \,,
\end{eqnarray}
we cancel all $\gamma^i$ terms from the action and its variation,  leaving
\begin{eqnarray}
	S_{\rm A} + S_{{\rm ct}, A} = L^3  \int d^4x \,  \alpha_i \beta^i\,, \qquad
	\delta S_{\rm A} + \delta S_{{\rm ct}, A} = 2L^3  \int d^4x \, \delta \alpha_i \beta^i \,.
\end{eqnarray}
These are now finite. The total action will be stationary on the boundary as well as in the bulk if we take $\delta \alpha_i(\vec{x}, t) = 0$, which identifies the compatible boundary condition as
\begin{eqnarray}
	\alpha_i(\vec{x}, t) = {\rm fixed} \,,
\end{eqnarray}
and thus $\alpha_i(\vec{x}, t)$ should be thought of as the source in the dual field theory. The operator it is a source for is then, 
\begin{eqnarray}
\label{ACurrent}
	\langle J^i(\vec{x}, t) \rangle \equiv {\delta S_{\rm tot} \over \delta \alpha_i(\vec{x}, t)} = 2L^3  \beta^i(\vec{x}, t) \,.
\end{eqnarray}
We have neglected an overall normalization for $S$ involving the five-dimensional Newton constant, but in principle this combines with $L^3$ to give a constant proportional to $N^2$, with $N$ the rank of the field theory gauge group. Thanks to (\ref{BetaDivergence}), we see this current is conserved, and thus we identify the  operator dual to $A_\mu$ with this boundary condition as a conserved global $U(1)$ symmetry current $J^i \sim \beta^i$, with non-dynamical background gauge field source $\alpha_i$. This is the regular boundary condition.

Let us make a few comments. We notice that the answer we get is the same as if we ignored the $\gamma_i \log r/r^2$ term, and also ignored the counterterm. However $\gamma_i$ will be important for our comparison to the magnetic dual $B_{\mu\nu}$ field, and we would like to be careful about what boundary conditions we are imposing by being specific about our counterterms, so we have included it.

We also note that while the coefficient of the $\log R$ term in (\ref{ECoefficient}) is fixed to cancel the divergence, we can in principle make the finite part whatever we want. A different choice for the finite part is equivalent to adding the counterterm
\begin{eqnarray}
\label{MaxwellCT}
	L^3 \int d^4x  \left(- {1 \over 4 e^2} f_{ij}f^{ij} \right)\,,
\end{eqnarray}
involving the background gauge field $\alpha_i$ only. This would preserve the boundary condition $\alpha(\vec{x},t)$ fixed, but shift the current,
\begin{eqnarray}
	\langle J^i(\vec{x}, t) \rangle \to  2L^3  \left( \beta^i(\vec{x}, t) + {1 \over 2e^2} \partial_j f^{ji}\right)\,,
\end{eqnarray}
which is still conserved; this shift has the form of an ``improvement" term that doesn't change the total charge. Adding such a counterterm involving a non-dynamical background field is not something we have need to do, but as we will see, the corresponding freedom  to add counterterms for the case of $B_{\mu\nu}$ will be more interesting to us.

We could also consider an alternate boundary condition for $A_\mu$, where $\beta_i(\vec{x},t)$ would be fixed while $\alpha_i(\vec{x},t)$ would be the fluctuating  dynamical degree of freedom, a dimension 1 field with a standard abelian gauge invariance: that is, the dynamical degree of freedom is a $U(1)$ gauge field. It has been noticed before, especially in the context of AdS$_4$, that alternate boundary conditions for a vector field lead to a gauged, rather than a global, $U(1)$ symmetry in the dual theory \cite{Witten:2003ya, Leigh:2003gk, Leigh:2003ez, Petkou:2004nu, Yee:2004ju, Marolf:2006nd}. We will not pursue the appropriate counterterms for this boundary condition directly in the $A_\mu$ language (for further discussion, see \cite{Marolf:2006nd}), but instead will show how the dynamics arises as the regular boundary condition for the magnetic dual 2-form field $B_{\mu\nu}$.

\subsection{The two-form $B_{\mu\nu}$ in AdS$_5$ gauges the global $U(1)$ dual to $A_\mu$}

Instead of (\ref{AAction}), consider the action for a two-form field $B_{\mu\nu}$ in AdS$_5$,
\begin{equation}
\label{BAction}
    S_B =  \int d^5x \sqrt{-g}\left( - {1 \over 12} H_{\mu\nu\rho} H^{\mu\nu\rho} \right) \,,
\end{equation}
with standard field strength $H_{\mu\nu\rho} = \partial_\mu B_{\nu\rho} + \partial_\nu B_{\rho \mu} + \partial_\rho B_{\mu\nu}$.\footnote{Note that a potential Chern-Simons term of the form $\int B\wedge H$ is a total derivative in the bulk. Chern-Simons terms containing two different 2-form fields are relevant in compactification of type IIB string theory on five-manifolds \cite{Witten:1998wy, Belov:2004ht} and are also examined more recently in \cite{Hofman:2017vwr}, but are not considered here.} The equation of motion is $\partial_\gamma (\sqrt{-g} H^{\alpha\beta\gamma}) = 0$, becoming
\begin{equation}
\eta^{ij} \partial_i H_{jk r} =0 \,, \quad \quad
\left(	\partial_r + {1 \over r} \right) H_{r ij} + {L^4\over r^4} \eta^{kl} \partial_k H_{lij} = 0 \,.\end{equation}
Choosing a gauge with $B_{ir} = 0$, we have the solution 
\begin{equation}
\label{BSoln}
    B_{ij} (\vec{x}, t,r) =  \tilde\alpha_{ij} (\vec{x}, t) L^2+   \tilde\beta_{ij}(\vec{x}, t) L^2\log r + \ldots \,,
\end{equation}
where the subleading terms do not contribute to the action or its variation in the large-$r$ limit. Analogously to the case of $A_\mu$, $\tilde\alpha_{ij} (\vec{x}, t)$ inherits the residual gauge invariance of $B_{\mu\nu}$,
\begin{eqnarray}
\label{BGaugeInvariance}
	\tilde\alpha_{ij} (\vec{x}, t) \to \tilde\alpha_{ij} (\vec{x}, t) + \partial_i \Lambda_j (\vec{x}, t) - \partial_j \Lambda_i (\vec{x}, t) \,,
\end{eqnarray}
and the equations of motion require $\tilde\beta_{ij} (\vec{x}, t)$ to be divergence-free,
\begin{eqnarray}
\label{BetaDivergence2}
	\partial_i \tilde\beta^{ij} =0 \,.
\end{eqnarray}
Scaling tells us that both $\tilde\alpha_{ij}$  and $\tilde\beta_{ij}$ have scaling dimension 2, and the fact that these dimensions coincide leads to the logarithm distinguishing the independent solutions in (\ref{BSoln}). 
As noted by \cite{Grozdanov:2017kyl, Hofman:2017vwr}, this logarithmic behavior leads to boundary conditions incompatible with conformal invariance; we will have more to say about this shortly.

The action and variation of the action evaluated on the solution are
\begin{eqnarray}
S_B = - {L^3 \over 4} \int d^4x \left( \tilde\alpha_{ij} +  \tilde\beta_{ij} \log R\right) \tilde\beta^{ij} \,, \qquad
	\delta S_B = - {L^3 \over 2} \int d^4x \left( \delta \tilde\alpha_{ij} + \delta \tilde\beta_{ij} \log R\right) \tilde\beta^{ij} \,, 
\end{eqnarray}
which are again divergent, thanks to the $\log R$ term. We can add a counterterm analogous to (\ref{ACounterterm}),
\begin{eqnarray}
\label{BCounterterm}
	S_{{\rm ct}, B} = L \int d^4x \sqrt{-h} \left( - {1 \over 12 \tilde{e}^2(R)} H_{\mu\nu\rho} H^{\mu\nu\rho} \right)\,.
\end{eqnarray}
and we obtain
\begin{eqnarray}
\label{BActionwithCT}
S_B + S_{{\rm ct}, B}  &=& - {L^3 \over 4} \int d^4x \left[ \tilde\alpha_{ij} + \left({1 \over \tilde{e}^2(R)} + \log R \right) \tilde\beta_{ij} \right] \tilde\beta^{ij} \,, \\
	\delta S_B + \delta S_{{\rm ct}, B} &=& - {L^3 \over 2} \int d^4x \left[ \delta \tilde\alpha_{ij} +\left({1 \over \tilde{e}^2(R)} + \log R \right) \delta \tilde\beta_{ij} \right] \tilde\beta^{ij} \,.  \nonumber
\end{eqnarray}
To cancel the divergence, we wish to have
\begin{eqnarray}
\label{ETildeCoefficient}
{1 \over \tilde{e}^2} =	{1 \over \tilde{e}^2(R)} + \log R \,,
\end{eqnarray}
for some $R$-independent constant $1/\tilde{e}^2$. For the  choice $1/\tilde{e}^2 = 0$ we obtain simply the boundary condition 
\begin{eqnarray}
	\tilde\alpha_{ij}(\vec{x}, t) = {\rm fixed} \,,
	\end{eqnarray}
and then $\tilde\alpha_{ij}(\vec{x}, t)$ is the source, a background 2-form gauge field for the 2-form current,
\begin{eqnarray}
\label{TwoFormCurrent}
	\langle J^{ij}(\vec{x}, t) \rangle \equiv {\delta S_{\rm tot} \over \delta \tilde\alpha_{ij}(\vec{x}, t)}= - {L^3 \over 2} \tilde\beta^{ij}(\vec{x}, t) \,,
\end{eqnarray}
which is conserved thanks to (\ref{BetaDivergence2}). In general we obtain the family of boundary conditions,
\begin{eqnarray}
\label{FamilyofBCs}
	\tilde\alpha_{ij}(\vec{x}, t) + {1 \over \tilde{e}^2} \, \tilde\beta_{ij}(\vec{x}, t) = {\rm fixed} \,,
\end{eqnarray}
while the current (\ref{TwoFormCurrent}) does not change. Since this family includes the case where $\tilde\alpha_{ij}(\vec{x},t)$ alone is fixed, but does not include a case where $\tilde\beta_{ij}(\vec{x},t)$ alone is fixed, we may think of it as a family of regular boundary conditions for $B_{\mu\nu}$.

Boundary conditions relating the $\alpha$ and $\beta$ modes to each other as in (\ref{FamilyofBCs}) are associated to the addition of multi-trace operators to the theory. Indeed, for $1/\tilde{e}^2 \neq 0$, according to (\ref{BActionwithCT}) we have added to the theory the boundary term,
\begin{eqnarray}
\label{DoubleTrace}
	L^3 \int d^4x \left(- {1 \over 4 \tilde{e}^2 } \, \tilde\beta_{ij} \tilde\beta^{ij} \right)\,.
\end{eqnarray}
Since $\tilde\beta_{ij} \sim J_{ij}$ is a single-trace operator (that is, it is gauge-invariant from the point of view of gauge symmetries of the field theory), this counterterm corresponds to the addition of a double-trace operator $JJ$. 
This double-trace operator was noted by \cite{Grozdanov:2017kyl, Hofman:2017vwr}. We will elaborate on the physics of this momentarily, after we describe the relationship between $A_\mu$ and $B_{\mu\nu}$ if we think of them electric and magnetic duals of each other.

If we declare the field strength of $B_{\mu\nu}$ to be dual to a field strength for $A_\mu$,
\begin{eqnarray}
\label{EMDual}
	H_{\mu\nu\rho} = {1 \over 2} \epsilon_{\mu\nu\rho}^{\phantom{\mu}\phantom{\nu}\phantom{\rho}\alpha\beta} F_{\alpha \beta} \,,
\end{eqnarray}
and insert the solutions (\ref{ASoln}) and (\ref{BSoln}), we find the correspondence\footnote{In our conventions the Levi-Civita symbols (with elements $+1$, $-1$ and $0$) are $\tilde\epsilon_{\mu\nu\rho\sigma\alpha}$ and $\tilde\epsilon_{ijkl}$, with 
$\tilde\epsilon_{\mu\nu\rho\sigma r} = \tilde\epsilon_{ijkl}$, and the corresponding Levi-Civita tensors are as usual $\epsilon_{\mu\nu\rho\sigma\alpha} = \sqrt{-g} \,\tilde\epsilon_{\mu\nu\rho\sigma\alpha}$ and $\epsilon_{ijkl} = \sqrt{-h}\,\tilde\epsilon_{ijkl}$.}
\begin{eqnarray}
\label{BetaTildeFromF}
	\tilde\beta_{ij} &=& {1 \over 2} \tilde\epsilon_{ijkl} f^{kl} \,, \\
	\tilde{f}_{ijk} &=& \tilde\epsilon_{ijkl} \Big( 2 \beta^l - {1 \over 2} \partial_m f^{ml} \Big) \,,
	\label{FTildeFromBeta}
\end{eqnarray}
where $\tilde{f}_{ijk} \equiv \partial_i \tilde\alpha_{jk}+ \partial_j \tilde\alpha_{ki} +\partial_k \tilde\alpha_{ij}$.
We note that including the $\gamma_i$ term in (\ref{ASoln}) was essential for matching the $\log r$ term in (\ref{BSoln}). The relations (\ref{BetaTildeFromF}), (\ref{FTildeFromBeta}) are gauge invariant under both gauge invariances (\ref{AGaugeInvariance}) and (\ref{BGaugeInvariance}), with $\alpha_i$ and $\tilde\alpha_{ij}$ appearing only through their field strengths. The inverse formulas are
\begin{eqnarray}
\label{AlphaFromH}
	f_{ij} &=& {1 \over 2} \tilde\epsilon_{ijkl} \tilde\beta^{kl} \,, \\
	\beta_i &=& {1 \over 24} \tilde\epsilon_{ijkl} \Big( 2 \tilde{f}^{jkl} + \tilde{g}^{jkl} \Big) \,,
	\label{BetaFromH}
\end{eqnarray}
with $\tilde{g}_{ijk} \equiv \partial_i \tilde\beta_{jk}+ \partial_j \tilde\beta_{ki} +\partial_k \tilde\beta_{ij}$.

From this we can see that regular boundary conditions for one field, by the time they are translated through the electric/magnetic duality (\ref{EMDual}), become alternate boundary conditions for the other field.  The regular boundary conditions we found for $A_\mu$ were $\delta \alpha_i(\vec{x},t ) =0$, which through (\ref{BetaTildeFromF}) imply $\delta \tilde\beta_{ij}(\vec{x},t ) =0$, which are alternate boundary conditions for $B_{\mu\nu}$,
\begin{eqnarray}
\label{BoundaryConds1}
	\alpha_i(\vec{x},t ) = {\rm fixed}\quad \rightarrow \quad \tilde\beta_{ij}(\vec{x},t ) = {\rm fixed} \,.
\end{eqnarray}
Meanwhile, the family of regular boundary conditions for $B_{\mu\nu}$ (\ref{FamilyofBCs}), which we could always think of as allowing $\tilde\beta_{ij}(\vec{x},t)$ to fluctuate while $\tilde\alpha_{ij}(\vec{x},t)$ is determined, maps to
\begin{eqnarray}
\label{BoundaryConds2}
	\tilde\alpha_{ij}(\vec{x}, t) + {1 \over \tilde{e}^2} \, \tilde\beta_{ij}(\vec{x}, t) = {\rm fixed} \quad \rightarrow \quad 2 \beta^i(\vec{x}, t) + \left({1 \over \tilde{e}^2} - {1 \over 2} \right)\partial_j f^{ji}(\vec{x}, t) = {\rm fixed} \,,
\end{eqnarray}
a family of boundary conditions for $A_\mu$ that we may think of as alternate conditions, since they may always be thought of as $\alpha_i(\vec{x},t)$ being free and $\beta_i(\vec{x},t)$ being determined; for the special case $\tilde{e}^2 = 2$ the conditions are precisely $\beta_i(\vec{x},t)$ fixed.

It is interesting how choosing between expressing identical bulk dynamics as $A_\mu$ or as $B_{\mu\nu}$ changes the boundary conditions. Indeed, the duality relation (\ref{EMDual}) implies
\begin{eqnarray}
	- {1 \over 12} H_{\mu\nu\rho} H^{\mu\nu\rho} =  {1 \over 4} F_{\mu\nu} F^{\mu\nu} \,,
\end{eqnarray}
and thus up to the overall sign, both the bulk Maxwell action (\ref{AAction}) and the counterterm  (\ref{ACounterterm}) for the $A$ field are equal to the bulk action (\ref{BAction}) and counterterm (\ref{BCounterterm}) for the $B$ field, with $1/e^2(R) \leftrightarrow 1/\tilde{e}^2(R)$, and indeed the same choice of divergent part in $1/e^2(R)$ (\ref{ECoefficient}) and  $1/\tilde{e}^2(R)$ (\ref{ETildeCoefficient}) renders both finite. Yet by the time we expand the fields near the boundary, the boundary conditions come out different; the identical choice $1/e^2(R)  = 1/\tilde{e}^2(R) = - \log R + 1/2$ leads to regular boundary conditions $\delta \alpha_i(\vec{x},t) = 0$ for $A_\mu$, but boundary conditions on $B_{\mu\nu}$ that when translated into $A_\mu$ variables are the alternate boundary conditions $\delta \beta_i(\vec{x},t) = 0$. In some sense, electromagnetic duality does not commute with the near-boundary limit. This is likely related to the fact that the gauge invariances for $A_\mu$ and $B_{\mu\nu}$ are different, and while absent the boundary these would be physically irrelevant redundancies, in the presence of the boundary each can induce a ``long-range gauge symmetry" which is physical and controls the dual global symmetry \cite{Harlow:2018tng}.

\begin{figure}
    \centering
    \includegraphics[width=1\linewidth]{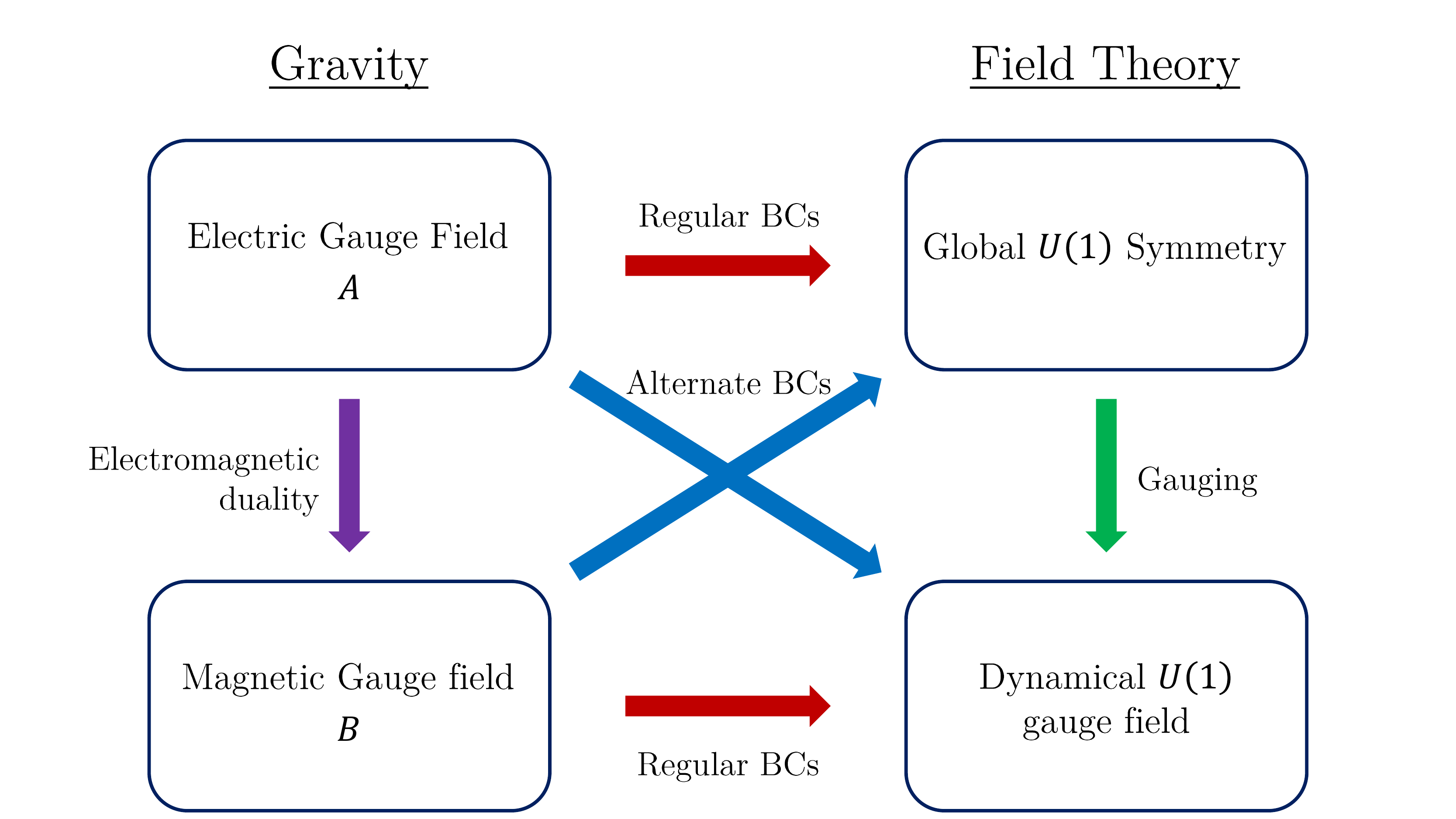}
    \caption{A flowchart depicting the relationships between electromagnetic duality, boundary conditions and gauging described in this section.}
    \label{fig:flowchart1}
\end{figure}

Recall from our earlier remarks that the alternate boundary condition for $A_\mu$ corresponds to a dynamical $U(1)$ vector field. But we now see this physics of a dynamical $U(1)$ gauge field should also find realization in $B_{\mu\nu}$ with the family of regular boundary conditions, for which the dual dynamical operator is a conserved 2-form current $J^{ij}$ (\ref{TwoFormCurrent}). Indeed according to (\ref{BetaTildeFromF}), this operator is nothing more than the dual field strength associated with $\alpha_i(\vec{x},t)$:
\begin{eqnarray}
	\langle J^{ij} (\vec{x},t) \rangle \propto \tilde\epsilon^{ijkl} f_{kl}(\vec{x},t) \,.
\end{eqnarray}
This is entirely natural: given any $U(1)$ vector field $\alpha_i$, the associated dual field strength is conserved by the Bianchi identity. Thus in the $A_\mu$ and $B_{\mu\nu}$ perspectives we see the same physics in two different ways: in the former we see a dynamical $U(1)$ gauge field itself, while in the latter we see its associated global 2-form current. One useful thing about generalized symmetries is that they can describe aspects of a gauge theory without referring to the gauge field itself, and the physics of $B_{\mu\nu}$ is a realization of this phenomenon. The relationships between boundary conditions, electromagnetic duality and gauging the global symmetry of the field theory are summarized in figure~\ref{fig:flowchart1}.

We can also now understand the physical meaning of the multi-trace operator (\ref{DoubleTrace}). Translated into $A_\mu$ variables, it is nothing more than a Maxwell kinetic term for the gauge field $\alpha_i$,
\begin{eqnarray}
	L^3\int d^4x \left(- {1 \over 4 \tilde{e}^2 } \, f_{ij} f^{ij} \right)\,.
\end{eqnarray}
This is analogous to (\ref{MaxwellCT}), except with these boundary conditions $f_{ij}$ is dynamical, instead of a background field. From (\ref{ETildeCoefficient}), we can understand the running of the coupling, since $1/\tilde{e}^2$ is cutoff-independent,
\begin{eqnarray}
	R {d \over dR} \left({1 \over \tilde{e}^2(R)} + \log R \right) =0 \quad \to\quad R {d \over dR} {1 \over \tilde{e}^2} =-1 \,,
\end{eqnarray}
and we find an IR-free coupling, as expected for a $U(1)$ gauge field coupled to electric matter. This is also consistent with the emergence of a linearly dispersing mode at high energies in the finite-temperature hydrodynamics for this field studied in \cite{Grozdanov:2017kyl, Hofman:2017vwr}.

Thus the logarithmic behavior of the $B_{\mu\nu}$ field in AdS$_5$ has a simple explanation: the dual 2-form current is just the magnetic field strength for a dynamical $U(1)$ gauge field in the dual theory, and the double-trace operator built from the 2-form current is just the gauge field's Maxwell kinetic term, with logarithmically running coupling, as in QED. Passing from $A_\mu$ to $B_{\mu\nu}$ has gauged the global $U(1)$. Moreover, the fact that any free $B_{\mu\nu}$ can be understood in this way is the holographic version of the statement that any conserved 2-form current in a four dimensional CFT can be expressed as the dual field strength of a Maxwell field \cite{Cordova:2018cvg}.

\subsection{Electromagnetic duality and generalized symmetries in general dimension}

Before turning to a discussion of the gravity dual to 2-group symmetries in four-dimensional field theories, we discuss briefly the generalization of the duality between $A_\mu$ and $B_{\mu\nu}$ in AdS$_5$ and its relation to gauging a global symmetry to a general pair of electric/magnetic dual $p$-form fields in AdS$_{d+1}$.

For a quantum field theory in $d$ dimensions possessing a conserved $p$-form current $j_p$, gauging the symmetry produces a dynamical $p$-form gauge field $\alpha_p$. Given the field strength $f_{p+1} = d\alpha_p$, we can always define a new $d-p-1$-form current
\begin{eqnarray}
\label{CurrentFromFieldStrength}
	J_{d-p-1} \equiv * f_{p+1} \,,
\end{eqnarray}
which is automatically conserved due to the Bianchi identity (assuming no magnetic sources for $\alpha_p$). The classic example of this is in $d=3$, where an ordinary 1-form gauge field leads to a 1-form conserved current via $J_1 = * f_2$, or $J^\mu = (1/2) \epsilon^{\mu\nu\rho} f_{\nu\rho}$; in this paper so far we discussed the case $p=1$, $d=4$.

It is easy to see that for general $d$ and $p$ we can reproduce this on the gravity side. Let the $p$-form current be realized in the AdS$_{d+1}$ gravity theory by the $p$-form field $A_p$, with a Maxwell action. The solution for the gauge where $A_p$ has no nonzero term with an $r$ index is
\begin{eqnarray}
	A_p(\vec{x},t,r) = \alpha_p(\vec{x},t) L^p + \ldots + \beta_p(\vec{x},t) {L^{2d-3p} \over r^{d-2p} }+ \ldots \,,
\end{eqnarray}
with $d * \beta_p = 0$, where $*$ is the $d$-dimensional Hodge star. With regular boundary conditions, this is dual to a conserved current $\langle j_p(\vec{x},t)\rangle \propto \beta_p(\vec{x},t)$, with $\alpha_p(\vec{x},t)$ the background gauge field.

To gauge this symmetry, consider the magnetic dual field $B_{d-p-1}$ in AdS$_{d+1}$,
\begin{eqnarray}
	B_{d-p-1}(\vec{x},t,r) = \tilde\alpha_{d-p-1}(\vec{x},t) L^{d-p-1} + \ldots +\tilde\beta_{d-p-1}(\vec{x},t) { L^{3p + 3-d} \over r^{2p+2-d} }+ \ldots  \,,
\end{eqnarray} 
and we find that electromagnetic duality $dB_{d-p-1} = * dA_p$ gives us relations of the form
\begin{eqnarray}
\label{GeneralDuality}
	\tilde\beta_{d-p-1} \propto * f_{p+1}  \,, \quad \quad \tilde{f}_{d-p} \propto * \beta_p\,.
\end{eqnarray}
Thus once more the regular boundary conditions for $B_{d-p-1}$ are the alternate boundary conditions for $A_p$, with the dynamical conserved current dual to $B_{d-p-1}$ related to the field strength of the gauged original symmetry as described above,
\begin{eqnarray}
	\langle J_{d-p-1} \rangle \propto \tilde\beta_{d-p-1} \propto * f_{p+1} \,.
\end{eqnarray}
Thus for general $d$ and $p$, the conserved $(d-p-1)$-form current that comes from gauging a global $p$-form current can be realized by electric/magnetic duality in the bulk. The flowchart of figure~\ref{fig:flowchart1} still describes this more general case. Notice that which bulk field we call electric and which we call magnetic is a matter of convention, and both cases in the field theory column of figure~\ref{fig:flowchart1} can be thought of as containing both a global conserved current and a dynamical gauge field, where one determines the other as in (\ref{CurrentFromFieldStrength}).

For general $d$ and $p$ the equations (\ref{GeneralDuality}) are simpler than their $d=4$, $p=1$ counterparts (\ref{BetaTildeFromF}), (\ref{FTildeFromBeta}) due to the absence of the log in the solution for $B$; the lack of the log indicates that the coupling for the double-trace operator $J_{d-p-1} \,J_{d-p-1}$ is not marginal. The log will arise for even $d$ when $p = d/2-1$ and hence $d-p-1 = d/2$ (or vice versa) for which the ranks of $A$ and $B$ differ only by one; these cases will be analogous to the AdS$_5$ case studied earlier.

\section{Gauging 't Hooft anomalies and 2-group symmetries}
\label{ChernSimonsModfiedSec}

We saw in the previous section that to gauge a global $U(1)$ in the field theory, we switch from regular to alternate boundary conditions in the dual vector field $A_\mu$, or equivalently replace $A_\mu$ with the 2-form $B_{\mu\nu}$. Thus gauging the $U(1)$ becomes electromagnetic duality on the gravity side.

We can realize a nice example of this with a simple model discussed by C\'ordova, Dumitrescu and Intriligator \cite{Cordova:2018cvg}. This model starts as a field theory with $U(1) \times U(1)$ global symmetry, with a mixed 't Hooft anomaly. Gauging one of the $U(1)$s introduces a  conserved 2-form current, which combines with the remaining global $U(1)$ into a symmetry structure called a 2-group. It is known that the ungauged theory should be modeled in the gravity dual by a pair of 1-form gauge fields, with a Chern-Simons coupling capturing the 't Hooft anomaly. Electromagnetic duality transforming one gauge field into a 2-form then produces exactly the correct modified field strength to generate a dual field theory with 2-group global symmetry, as we shall see.

\subsection{The $U(1) \times U(1)$ model}

Here we briefly review the method presented by C\'ordova et al. for obtaining a 2-group symmetry structure from a mixed 't Hooft anomaly in \cite{Cordova:2018cvg}. Consider a four-dimensional quantum field theory manifesting $U(1)_A \times U(1)_C$ global symmetry with background 1-form gauge fields $A$ and $C$, with field strengths $F = dA$ and $G = dC$. Further suppose the theory has a mixed 't Hooft anomaly represented by the anomaly polynomial
\begin{equation}
    I^{(6)} = \kappa\, F \wedge F \wedge G\,, \label{eq:2grp_rev_I6}
\end{equation}
where $\kappa$ is a constant.
If $U(1)_A$ is gauged, this leads to an ordinary Adler-Bell-Jackiw anomaly for the global $U(1)_C$, $\partial^\mu J^C_\mu \propto \kappa \epsilon^{\mu\nu\rho\sigma}F_{\mu\nu} F_{\rho\sigma}$. Instead, consider the case where we gauge $U(1)_C$ while $U(1)_A$ remains global. Respecting the gauged $U(1)_C$ then requires we choose our counterterms such that the anomaly descent forms encoded in $I^{(6)} = d I^{(5)}$, $\delta I^{(5)} = d I^{(4)}$ are written as
\begin{equation}
I^{(5)} = 	\kappa\, A \wedge F \wedge G \,, \quad \quad
I^{(4)} = 	\kappa \, \lambda_A \, F \wedge G \,,
\end{equation}
where $\delta A = d \lambda_A$, and thus the variation of the effective action is
\begin{equation}
\label{ApparentAnomaly}
	\delta S = \int I^{(4)} = \kappa \int   \lambda_A \, F \wedge G \,. 
\end{equation}
This nontrivial variation of $U(1)_A$ does not have the form of an ordinary anomaly because the right-hand-side includes a dynamical gauge field alongside a background field. However, the gauging of $U(1)_C$ introduces a new 1-form generalized global symmetry, with 2-form current
\begin{equation}
    J_m = *G\,.
\end{equation}
This couples to a background 2-form gauge field $B$ via
\begin{equation}
    S_J = \int B \wedge G.
\end{equation}
The variation of the action (\ref{ApparentAnomaly}) will be canceled if we allow $B$ to vary under transformations of $A$ in addition to its own natural transformation:
\begin{equation}
\label{TwoGroupTrans}
    \delta B = d\Lambda_B - \kappa \lambda_A F \,, \quad \quad \delta A = d\lambda_A\,,
\end{equation}
where $\Lambda_B$ is a 1-form gauge parameter.
With this combined transformation, the effective action as a function of the background fields $S[A,B]$ is invariant. This is known as a 2-group symmetry, and is denoted
\begin{equation}
    U(1)_A^{(0)} \times_\kappa U(1)_B^{(1)} \,, 
\end{equation}
where the superscripts indicate that $U(1)_A^{(0)}$ is an ordinary symmetry with a 0-form parameter $\lambda_A$ while $U(1)_B^{(1)}$ is a generalized symmetry with 1-form parameter $\Lambda_B$, and the presence of $\kappa$ indicates the nontrivial transformation of $B$ in (\ref{TwoGroupTrans}).\footnote{\cite{Cordova:2018cvg} actually writes $U(1)_A^{(0)} \times_{\hat\kappa} U(1)_B^{(1)} $ where $\hat\kappa \equiv 2 \pi \kappa$ is integer quantized.}
The invariant field strength $H$ for $B$ is modified by this 2-group symmetry structure:
\begin{equation}
\label{BackgroundFieldStrength}
    H = dB +\kappa  A\wedge F.
\end{equation}

\subsection{Gravity dual of the global $U(1) \times U(1)$ is a Chern-Simons theory} 

For the gravity dual of the global $U(1)_A \times U(1)_C$ field theory, we introduce two 1-form fields $A$ and $C$ in AdS$_5$, as described in section~\ref{OneFormTwoFormSec}. In addition, the presence of the 't Hooft anomaly is well-known to be described by a Chern-Simons term in the gravity theory \cite{Witten:1998qj} whose structure mimics the anomaly polynomial $I^{(5)} = \kappa \, A \wedge F \wedge G$. We thus have the bulk action
\begin{eqnarray}
\label{CSAction}
    S &=&  \int \left[ - {1 \over 2} F \wedge *F - {1 \over 2} G \wedge * G + \kappa \, A \wedge F \wedge G \right]\\
   &=& \int d^5x \, \sqrt{-g} \left[ - \frac{1}{4} F_{\mu\nu} F^{\mu\nu}  -\frac{1}{4} G_{\mu\nu} G^{\mu\nu}+ \frac{\kappa}{4} \epsilon^{\alpha\mu\nu\rho\sigma} A_\alpha F_{\mu\nu} G_{\rho\sigma} \right]\,.  \nonumber
\end{eqnarray}
The bulk equations of motion are
\begin{eqnarray}
\label{CSEOMsForm}
	d*F = 2 \kappa F \wedge G \,, \quad  \quad d*G = \kappa F \wedge F\,,
\end{eqnarray}
or equivalently,
\begin{eqnarray}
\label{CSEOMs}
	\partial_\nu \sqrt{-g} F^{\mu\nu} = {\kappa \over 2} \tilde\epsilon^{\mu\rho\sigma\alpha\beta} F_{\rho \sigma} G_{\alpha\beta}\,, \quad \quad \partial_\nu \sqrt{-g}\, G^{\mu\nu} = {\kappa \over 4} \tilde\epsilon^{\mu\rho\sigma\alpha\beta} F_{\rho \sigma} F_{\alpha\beta}\,,
\end{eqnarray}
while the Bianchi identities are simply
\begin{eqnarray}
\label{CSBianchis}
	dF =0 \,, \quad \quad dG =0 \,.
\end{eqnarray}
The solutions to the equations of motion (\ref{CSEOMs}) still take the form (\ref{ASoln}),
\begin{eqnarray}
	A_i &=& \alpha_{A, i}(\vec{x},t) L + \ldots + \beta_{A, i}(\vec{x},t){ L^5 \over r^2} + \ldots\\
	C_i &=& \alpha_{C, i}(\vec{x},t) L + \ldots + \beta_{C, i}(\vec{x},t){ L^5 \over r^2} + \ldots \,.
\end{eqnarray}
We have suppressed the $\gamma_i$ terms, but they lead to divergences canceled by counterterms just as in section~\ref{OneFormTwoFormSec}. The only effect of the nonlinear Chern-Simons terms in (\ref{CSEOMs}) to the leading $\alpha$ and $\beta$ coefficients is the modification of the divergences,
\begin{eqnarray}
\label{CSDivergences}
	\partial_i \beta^i_A =  - {\kappa \over 4} \tilde\epsilon^{ijkl} f_{ij} g_{jk} \,, \quad \quad 
	\partial_i \beta^i_C =- {\kappa \over 8} \tilde\epsilon^{ijkl} f_{ij} f_{jk} \,,
\end{eqnarray}
where $f = d \alpha_A$, $g = d\alpha_C$.

Including the counterterms as before, the variation of the action is now
\begin{eqnarray}
	\delta S =2L^3 \int d^4 x   \left[\delta\alpha_{A,i} \left( \beta_A^i + {\kappa \over 4} \tilde\epsilon^{ijkl} \alpha_{A,j} \, g_{kl} \right)+ \delta \alpha_{C,i} \left( \beta_C^i + {\kappa \over 4} \tilde\epsilon^{ijkl} \alpha_{A,j} \, f_{kl}\right)\right]  \,,
\end{eqnarray}
and thus we find the currents
\begin{eqnarray}
\label{CSCurrent1}
	\langle J_A^i \rangle &=& 2 L^3 \left( \beta_A^i + {\kappa \over 4} \tilde\epsilon^{ijkl} \alpha_{A,j} g_{kl}\right) \\
	\langle J_C^i \rangle &=& 2 L^3 \left( \beta_C^i + {\kappa \over 4} \tilde\epsilon^{ijkl} \alpha_{A,j} f_{kl}\right) \,.
	\label{CSCurrent2}
\end{eqnarray}
A $U(1)_A \times U(1)_C$ transformation can be written
\begin{eqnarray}
	\delta \alpha_{A,i}(\vec{x},t) = \partial_i \lambda_A(\vec{x},t) \,, \quad \quad 
	\delta \alpha_{C,i}(\vec{x},t) = \partial_i \lambda_C(\vec{x},t) \,,
\end{eqnarray}
and then the variation of the action is
\begin{eqnarray}
	\delta S = -  \int d^4x \left(\lambda_A \, \partial_i \langle J_A^i\rangle + \lambda_C \, \partial_i \langle J_C^i\rangle  \right) \,,
\end{eqnarray}
where the divergences of the currents evaluate to
\begin{eqnarray}
\label{DivergenceJA}
	\partial_i \langle J_A^i \rangle &=& -{ L^3 \kappa \over 4}\,  \tilde\epsilon^{ijkl} f_{ij}g_{kl} \,, \\
	\partial_i \langle J_C^i \rangle &=& 0 \,.
	\label{DivergenceJC}
\end{eqnarray}
As expected, the action is invariant under $U(1)_C$ but transforms like (\ref{ApparentAnomaly}) under $U(1)_A$, realizing the 't Hooft anomaly. At this point, neither $U(1)$ is gauged and there is no issue with this transformation. 

The effects of an 't Hooft anomaly can be rearranged by adding local counterterms; to accomplish this in the gravity theory, we could add boundary counterterms of the form $\int A \wedge C \wedge F$, equivalent to integrating the Chern-Simons action by parts. This would leave (\ref{CSDivergences}) unchanged but change (\ref{CSCurrent1}), (\ref{CSCurrent2}) and hence reshuffle the right-hand-sides of (\ref{DivergenceJA}), (\ref{DivergenceJC}); for example we could remove the divergence of $J_A^i$ at the expense of giving $J_C^i$ an ordinary Adler-Bell-Jackiw anomaly proportional to $\tilde\epsilon^{ijkl} f_{ij}f_{kl}$. However to gauge $U(1)_C$, as we will do next, we need its current to obey (\ref{DivergenceJC}), so we must leave the action as is. 

\subsection{Gravity dual of 2-group structure from electromagnetic duality}

From our previous discussion, we have learned that replacing a 1-form field in AdS$_5$ with the magnetic dual 2-form is equivalent to gauging the formerly global $U(1)$ symmetry. Thus, we might expect that by appropriately dualizing $C$ in the $U(1)_A \times U(1)_C$ Chern-Simons action (\ref{CSAction}) into a 2-form $B$, we would arrive at a gravity theory with 2-group symmetry in the field theory dual. Indeed we show this is the case.

We begin by defining a three-form field strength $H$ dual to the two-form $G$,
\begin{eqnarray}
	H \equiv * G \,,
\end{eqnarray}
which implies $-G = * H$, while retaining $F = dA$ as before.
The equations of motion (\ref{CSEOMsForm}) and Bianchi identities (\ref{CSBianchis}) of the Chern-Simons theory then become
\begin{eqnarray}
\label{ModifiedFSEqns1}
	d*F = -2 \kappa F \wedge *H \,, \quad  \quad dH = \kappa F \wedge F\,,
\end{eqnarray}
and
\begin{eqnarray}
\label{ModifiedFSEqns2}
	dF = 0 \,, \quad \quad d*H = 0 \,.
\end{eqnarray}
To generate this nontrivial Bianchi identity for $H$, we assume the modified field strength form
\begin{eqnarray}
\label{HFieldStrength}	
H \equiv dB + \kappa A \wedge F \,,	
\end{eqnarray}
for a new dynamical two-form potential $B$. We still have $dF = 0$. One can then verify that the action
\begin{eqnarray}
S = \int \left( - {1 \over 2} H \wedge *H - {1 \over 2} F \wedge * F \right) \,, \label{eq:dualAction}
\end{eqnarray}
gives rise to the other two equations in (\ref{ModifiedFSEqns1}), (\ref{ModifiedFSEqns2}) as equations of motion.

The field strengths $H$ (\ref{HFieldStrength}) and $F$ are invariant under the gauge transformations
\begin{eqnarray}
	\delta B = d \Lambda_B - \kappa \lambda_A F  \,, \quad \quad \delta A = d \lambda_A\,, \label{eq:ABgaugeTrans}
\end{eqnarray}
where $\Lambda_B$ is the natural 1-form gauge parameter for the 2-form $B$. This set of transformations is exactly the gauged version of the 2-group transformation (\ref{TwoGroupTrans}), and the modified field strength (\ref{HFieldStrength}) is the same form as the 2-group background field strength (\ref{BackgroundFieldStrength}). Indeed, due to this correspondence C\'ordova et al.~proposed the modified field strength (\ref{HFieldStrength}) as the appropriate gravity dual form for a 2-group symmetry \cite{Cordova:2018cvg}. We now see that it arises naturally from dualizing the Chern-Simons action describing the ungauged theory via electromagnetic duality.

Let us now verify that the 2-group symmetry arises in the near-boundary analysis in detail. The equations of motion in component form are
\begin{eqnarray}
\label{ModfiedEOMsComponents}
	\partial_\rho \sqrt{-g} H^{\mu\nu\rho} =0 \,, \qquad 
	\partial_\nu \sqrt{-g} F^{\mu\nu} = - \kappa \sqrt{-g} H^{\mu\nu\rho} F_{\nu\rho} \,.
\end{eqnarray}
As in section~\ref{OneFormTwoFormSec}, we can use gauge freedom to set $A_r = B_{ir} = 0$,
and the fields then have the form\footnote{Note that in section~\ref{OneFormTwoFormSec}, we discussed the fields $A$ and $B$ being dual to each other; in this section, to match the notation of \cite{Cordova:2018cvg}, $B$ is dual to the  field $C$, while $A$ is independent.}
\begin{eqnarray}
\label{ModifiedA}
	A_i(\vec{x},t, r) &=& \alpha_i(\vec{x},t) L + \ldots + \beta_i(\vec{x},t) {L^5 \over r^2} + \ldots \\
	B_{ij} (\vec{x}, t,r) &=&  \tilde\alpha_{ij} (\vec{x}, t)L^2  +  \tilde\beta_{ij}(\vec{x}, t) L^2 \log r + \ldots \,.
	\label{ModifiedB}
\end{eqnarray}
The equations of motion (\ref{ModfiedEOMsComponents}) with the modified field strength do not modify the divergence of $\tilde\beta_{ij}$ from (\ref{BetaDivergence2}),
\begin{equation}
\label{ModifiedBetaTildeDivergence}
\partial_i \tilde\beta^{ij} = 0\,,	
\end{equation}
and their only modification to the modes appearing in (\ref{ModifiedA}), (\ref{ModifiedB}) is a nonzero divergence of $\beta_i$,
\begin{equation}
\label{ModifiedBetaDivergence}
    \partial_i \beta^i = {\kappa \over 2} f_{ij} \tilde\beta^{ij}.
\end{equation}
The action and its variation evaluated on these solutions reduce to divergent boundary expressions that look like the free cases discussed in section~\ref{OneFormTwoFormSec}, with one additional finite term. Choosing our counterterms as in section~\ref{OneFormTwoFormSec} to remove the divergences as well as extra finite parts, we find for the variation of the action
\begin{equation}
    \delta S = L^3 \int d^4x \left[ -\frac{1}{2} \delta \tilde\alpha_{ij}  \tilde\beta^{ij}+ \delta\alpha_i  (2 \, \beta^i - \kappa \, \alpha_j  \tilde\beta^{ij}) \right]. \label{eq:varDualAction}
\end{equation}
From this variation, we find two currents in the dual field theory:
\begin{align}
    \langle J^i_A \rangle &= L^3 \left( 2\beta^i  -\kappa \alpha_j \tilde\beta^{ij} \right)\,, \\
    \langle J^{ij}_B \rangle &= -\frac{L^3}{2} \tilde\beta^{ij}.
\end{align}
We note that the 2-form current is conserved by (\ref{ModifiedBetaTildeDivergence}), but the 1-form current is no longer conserved due to the nonzero divergence of $\beta_i$ (\ref{ModifiedBetaDivergence}):
\begin{eqnarray}
    \partial_i \langle J^i_A \rangle &=& {L^3 \kappa \over 2} f_{ij} \tilde\beta^{ij} = - \kappa f_{ij} \langle J^{ij}_B \rangle\,,\\
    \partial_i \langle J^{ij}_B \rangle &=& 0 \,.
\end{eqnarray}
The nonzero divergence for $J^i_A$ is the analog of the 't Hooft anomaly (\ref{DivergenceJA}), with the dynamical $\langle J^{ij}_B \rangle$ having replaced the background field $\tilde\epsilon^{ijkl} g_{kl}$ after the field $C$ was gauged.

\begin{figure}[h!]
    \centering
    \includegraphics[width=1.0\linewidth]{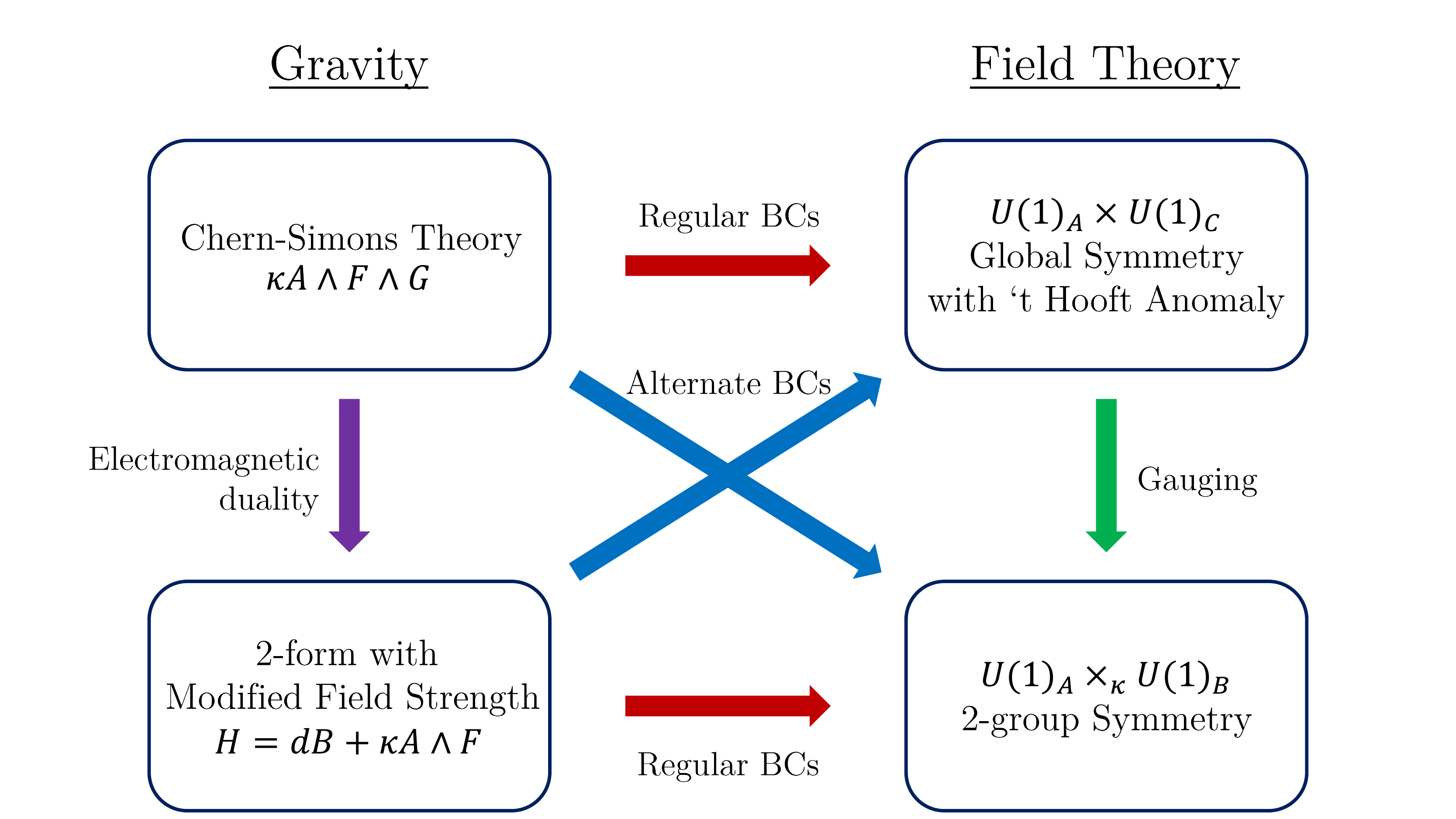}
    \caption{A diagram of how electromagnetic duality on a bulk Chern-Simons theory leads 
    to a theory with 2-group symmetry. In this case the blue arrows would represent alternate boundary conditions for the $B$ and $C$ fields, but still regular boundary conditions for $A$.}
    \label{fig:flowchart2}
\end{figure}

The conservation of $J^{ij}_B$ means that ordinary gauge transformations for the 2-form field $\delta \tilde\alpha_{ij} = \partial_i \Lambda_j - \partial_j \Lambda_i$ are symmetries of the effective action. Meanwhile if we perform a gauge transformation $\delta \alpha_i = \partial_i \lambda_A$ and some unspecified transformation $\delta \tilde\alpha_{ij}$, we find
\begin{eqnarray}
	\delta S =  \int d^4x \left[ -\frac{L^3}{2} \delta \tilde\alpha_{ij}  \tilde\beta^{ij}- \lambda_A \partial_i    \langle J^i_A \rangle \right] =- {L^3 \over 2} \int d^4x \left[ \delta \tilde\alpha_{ij}  \tilde\beta^{ij} + \kappa \,\lambda_A f_{ij} \tilde\beta^{ij} \right] \,. \nonumber
\end{eqnarray}
Thus $\delta \alpha_i = \partial_i \lambda_A$, $\delta \tilde\alpha_{ij} =0$ is not a symmetry of the effective action, but leads to a variation of the form (\ref{ApparentAnomaly}). Instead the action is invariant under the combined transformation of the 2-group form,
\begin{eqnarray}
	\delta \alpha_i = \partial_i \lambda_A\,, \quad \quad \delta \tilde\alpha_{ij} = - \kappa \,\lambda_A f_{ij} \,.
\end{eqnarray}
Thus the net gauge invariances of $\alpha_i$, $\tilde\alpha_{ij}$ are precisely that of the 2-group $U(1)_A^{(0)} \times_\kappa U(1)_B^{(1)}$,
\begin{align}
    \delta\alpha &= d\lambda_A \\
    \delta\tilde\alpha &= d\Lambda_B - \kappa \lambda_A f\,,  \nonumber
\end{align}
matching the transformations (\ref{eq:ABgaugeTrans}) required to keep the modified field strength tensor invariant. Thus indeed, we see gauging $C$ in (\ref{CSAction}) by performing electromagnetic duality to a 2-form gauge field produces a 2-group symmetry structure in the field theory dual. Figure~\ref{fig:flowchart2} summarizes the process used to obtain the 2-group symmetry.

Let us make a few comments. First, we can only dualize $C$ into $B$ despite the Chern-Simons term (\ref{CSAction}) because $C$ does not appear outside its field strength (while $A$ does). This is associated to the fact that we are preserving the $U(1)_C$ symmetry from the 't Hooft anomaly, as discussed in the last subsection; adding boundary counterterms to rearrange the bulk Chern-Simons term into a $C \wedge F \wedge F$ form would cause $U(1)_C$ to be violated and be ungaugeable, while electromagnetic duality on $C$ can no longer operate since it appears outside its field strength.

Second, for both the Chern-Simons theory (\ref{CSAction}) and the 2-group theory (\ref{eq:dualAction}), the non-linear terms in the equations of motion did not modify the cancellation of the divergences from the way it worked for the free field case, but only introduced new finite parts. This makes the identification of symmetry structures like t' Hooft anomalies and 2-group structure in the dual field theory possible without going through every detail of holographic renormalization, and is related to the scheme-independence of the symmetries. 

Finally, \cite{Cordova:2018cvg} notes that gauging the $U(1)_B^{(1)}$ symmetry returns us to the original ungauged $U(1)_A \times U(1)_C$ theory; this is naturally reflected for us in the fact that a second electromagnetic duality on $B$ returns us to the original electric variables.

\subsection{$n$-group symmetries in AdS/CFT in general dimension}

It is straightforward to generalize the 2-group structure discussed above. A more general $n$-group structure can be associated to background fields $B_n$, $A_p$, $C_{n-p}$ when the effective action is invariant under the transformations
\begin{eqnarray}
\label{NGroup}
	\delta A_p = d \lambda_{p-1} \,, \quad \quad \delta C_{n-p} = d \omega_{n-p-1}\,, \quad \quad  \delta B_n = d\Lambda_{n-1} - \kappa \lambda_{p-1} \wedge dC_{n-p} \,,
\end{eqnarray}
which respects the invariance of the field strengths
\begin{eqnarray}
	H_{n+1} = dB_n + \kappa A_p \wedge G_{n-p+1} \,, \quad \quad  G_{n-p+1} = d C_{n-p}\,, \quad \quad  F_{p+1} = d A_p \,.
\end{eqnarray}
It is straightforward to show this more general case can be implemented in AdS/CFT as well. For a $d$-dimensional field theory, consider the $(d+1)$-dimensional gravity action
\begin{eqnarray}
S = \int	d^{d+1}x \left( - {1 \over 2} H \wedge *H- {1 \over 2} G \wedge *G  - {1 \over 2} F \wedge *F \right) \,, 
\end{eqnarray}
which leads to equations of motion
\begin{eqnarray}
	d*F = (-1)^p \kappa\, G \wedge *H \,, \quad \quad d*G = (-1)^{np+1} \kappa\, F \wedge *H \,, \quad\quad d*H = 0\,,
\end{eqnarray}
implying solutions
\begin{align} \nonumber
	A_p(\vec{x},t,r) &= \alpha_p(\vec{x},t) \, L^p+ \ldots+ \ldots \beta_p(\vec{x},t) {L^{2d-3p} \over r^{d-2p} }\,, \\
	B_n(\vec{x},t,r) &= \tilde\alpha_n(\vec{x},t)\, L^n+ \ldots+ \ldots \tilde\beta_n(\vec{x},t) {L^{2d-3n} \over r^{d-2n} }\,, \\
	C_{n-p}(\vec{x},t,r) &= \hat\alpha_{n-p}(\vec{x},t)\, L^{n-p}+ \ldots+ \ldots \hat\beta_{n-p}(\vec{x},t) {L^{2d+3p-3n} \over r^{d+2p-2n} }
	 \,, \nonumber
\end{align}
obeying the constraints
\begin{align} \nonumber
	d*\beta_p &= (-1)^p \kappa {d-2n \over d-2p} (d \hat\alpha_{n-p} \wedge * \tilde\beta_n )\,, \\
	\label{GeneralBetaTildeConstraint}
	d*\hat\beta_{n-p} &= (-1)^{np+1} \kappa {d-2n \over d+2p-2n} (d \alpha_p \wedge * \tilde\beta_n )\,, \\ \nonumber
	d*\tilde\beta_n &=0\,.
\end{align}
Here we make the simplifying assumption that none of $p$, $n$ or $n-p$ equal $d/2$; otherwise logs would be present, but the calculation should be similar. We also assume $p \geq 1$. The variation of the action is then
\begin{eqnarray}
	\delta S = \int \left( \delta \alpha_p \wedge *J_A + \delta \tilde\alpha_n \wedge *J_B + \delta \hat\alpha_{n-p} \wedge *J_C \right) \,,
\end{eqnarray}
where the currents
\begin{align} \nonumber
	* J_A &\equiv (-1)^{d+1} L^{d-1} (d-2p) \, (* \beta_p) \,,\\
	* J_B &\equiv (-1)^{d+1} L^{d-1} (d-2n) \, (* \tilde\beta_n) \,,\label{JC} \\
	* J_C &\equiv (-1)^{d+1} L^{d-1} (d-2p-2n) \, (* \hat\beta_{n-p}) + (-1)^{np} \kappa L^{d-1} (d-2n) (\alpha_p \wedge * \tilde\beta_n)\,,  \nonumber
\end{align}
obey
\begin{align} \nonumber
	d*J_A &= (-1)^{d+p+1} \kappa L^{d-1} (d-2n) (d\hat\alpha_{n-p} \wedge * \tilde\beta_n )\,,\\
	d*J_B &=0  \,, \\
	d*J_C &=0 \,. \nonumber
\end{align}
The conservation of $J_C$ involves a nontrivial cancellation between its two terms using the middle term in (\ref{GeneralBetaTildeConstraint}), similar to the conservation of the current (\ref{DivergenceJC}) in the Chern-Simons action. The action is then invariant under
\begin{eqnarray}
	\delta \alpha_p = d \lambda_{p-1} \,, \quad \quad \delta \hat\alpha_{n-p} = d \omega_{n-p-1}\,, \quad \quad  \delta \tilde\alpha_n = d\Lambda_{n-1} - \kappa \lambda_{p-1} \wedge d\hat\alpha_{n-p} \,,
\end{eqnarray}
which is indeed the $n$-group structure (\ref{NGroup}).
For the case $p = n/2$ one may let $C$ and $A$ be the same field, as in the 2-group case that was our primary example; the calculation goes through analogously.

\section{Conclusions}
\label{ConclusionsSec}

In this paper, we have discussed how electromagnetic duality on bulk $p$-form gauge fields holographically realizes the fact that gauging any generalized global symmetry $j_p$ in a $d$-dimensional field theory leads to a novel generalized global symmetry $J_{d-p-1}$ from the dual field strength of the new gauge field. The transformation of an ungauged model with an 't Hooft anomaly to a gauged model with 2-group symmetry can be captured perfectly by this electromagnetic duality in the bulk. Special features like the logarithmic running of a coupling for a double-trace operator are associated to certain cases where the gauged symmetry has a classically marginal coupling.

There are a number of interesting directions for future study. While our primary example involved four-dimensional quantum field theory, there may be cases in particular other dimensions, for example six, where the holographic perspective may lead to new insights. Other features of theories with $n$-group symmetries, such as their spontaneous breaking, renormalization group flows, and anomalies, may be interesting to investigate in a holographic context. Finally, the nature of a gauge field in a theory with a boundary and the associated ``long-range gauge symmetry" is worth further investigation, and the nontrivial interplay between electromagnetic duality and the near-boundary limit in AdS/CFT may be able to shed more light on this.

\section*{Acknowledgments}

We are grateful to Ken Intriligator for comments on the draft, and to Sa\v{s}o Grozdanov, Greg Moore, Ioannis Papadimitriou and Urs Schreiber for discussions on related work. The authors are supported by the Department of Energy under grant DE-SC0010005. 

\bibliographystyle{utphys}
\bibliography{Generalized}

\providecommand{\href}[2]{#2}\begingroup\raggedright\begin{thebibliography}{10}

\bibitem{Gaiotto:2014kfa}
D.~Gaiotto, A.~Kapustin, N.~Seiberg, and B.~Willett, ``{Generalized Global
  Symmetries},'' \href{http://dx.doi.org/10.1007/JHEP02(2015)172}{{\em JHEP}
  {\bfseries 02} (2015) 172}, \href{http://arxiv.org/abs/1412.5148}{{\ttfamily
  arXiv:1412.5148 [hep-th]}}.

\bibitem{Cordova:2018cvg}
C.~C\'ordova, T.~T. Dumitrescu, and K.~Intriligator, ``{Exploring 2-Group
  Global Symmetries},'' \href{http://dx.doi.org/10.1007/JHEP02(2019)184}{{\em
  JHEP} {\bfseries 02} (2019) 184},
  \href{http://arxiv.org/abs/1802.04790}{{\ttfamily arXiv:1802.04790
  [hep-th]}}.

\bibitem{baez2003higher}
J.~C. Baez and A.~D. Lauda, ``Higher-dimensional algebra v: 2-groups,''
  \href{http://arxiv.org/abs/math/0307200}{{\ttfamily arXiv:math/0307200}}.

\bibitem{Baez:2004in}
J.~Baez and U.~Schreiber, ``{Higher gauge theory: 2-connections on
  2-bundles},'' \href{http://arxiv.org/abs/hep-th/0412325}{{\ttfamily
  arXiv:hep-th/0412325}}.

\bibitem{Baez:2005sn}
J.~C. Baez, D.~Stevenson, A.~S. Crans, and U.~Schreiber, ``{From loop groups to
  2-groups},'' \href{http://arxiv.org/abs/math/0504123}{{\ttfamily
  arXiv:math/0504123}}.

\bibitem{Baez:2005}
J.~Baez and U.~Schreiber, ``{Higher gauge theory},''
  \href{http://arxiv.org/abs/math/0511710}{{\ttfamily arXiv:math/0511710}}.

\bibitem{Sati:2008eg}
H.~Sati, U.~Schreiber, and J.~Stasheff,
  \href{http://dx.doi.org/10.1007/978-3-7643-8736-5_17}{``{$L_{\infty}$ algebra
  connections and applications to String- and Chern-Simons n-transport},''}
\newblock 2, 2008.
\newblock \href{http://arxiv.org/abs/0801.3480}{{\ttfamily arXiv:0801.3480
  [math.DG]}}.

\bibitem{Schreiber2008}
J.~Baez and U.~Schreiber, ``{Smooth Functors vs.~Differential Forms},''
  \href{http://arxiv.org/abs/0802.0663}{{\ttfamily arXiv:0802.0663}}.

\bibitem{Schreiber2008-2}
U.~Schreiber and K.~Waldorf, ``Connections on non-abelian gerbes and their
  holonomy,'' \href{http://arxiv.org/abs/0808.1923}{{\ttfamily arXiv:0808.1923
  [math.DG]}}.

\bibitem{Kapustin:2013uxa}
A.~Kapustin and R.~Thorngren, ``{Higher symmetry and gapped phases of gauge
  theories},'' \href{http://arxiv.org/abs/1309.4721}{{\ttfamily arXiv:1309.4721
  [hep-th]}}.

\bibitem{Tachikawa:2017gyf}
Y.~Tachikawa, ``{On gauging finite subgroups},''
  \href{http://dx.doi.org/10.21468/SciPostPhys.8.1.015}{{\em SciPost Phys.}
  {\bfseries 8} no.~1, (2020) 015},
  \href{http://arxiv.org/abs/1712.09542}{{\ttfamily arXiv:1712.09542
  [hep-th]}}.

\bibitem{Benini:2018reh}
F.~Benini, C.~C\'ordova, and P.-S. Hsin, ``{On 2-Group Global Symmetries and
  their Anomalies},'' \href{http://dx.doi.org/10.1007/JHEP03(2019)118}{{\em
  JHEP} {\bfseries 03} (2019) 118},
  \href{http://arxiv.org/abs/1803.09336}{{\ttfamily arXiv:1803.09336
  [hep-th]}}.

\bibitem{Cordova:2020tij}
C.~C\'ordova, T.~T. Dumitrescu, and K.~Intriligator, ``{2-Group Global
  Symmetries and Anomalies in Six-Dimensional Quantum Field Theories},''
  \href{http://arxiv.org/abs/2009.00138}{{\ttfamily arXiv:2009.00138
  [hep-th]}}.

\bibitem{DelZotto:2020sop}
M.~Del~Zotto and K.~Ohmori, ``{2-Group Symmetries of 6d Little String Theories
  and T-duality},'' \href{http://arxiv.org/abs/2009.03489}{{\ttfamily
  arXiv:2009.03489 [hep-th]}}.

\bibitem{Hidaka:2020iaz}
Y.~Hidaka, M.~Nitta, and R.~Yokokura, ``{Higher-form symmetries and 3-group in
  axion electrodynamics},''
  \href{http://dx.doi.org/10.1016/j.physletb.2020.135672}{{\em Phys. Lett. B}
  {\bfseries 808} (2020) 135672},
  \href{http://arxiv.org/abs/2006.12532}{{\ttfamily arXiv:2006.12532
  [hep-th]}}.

\bibitem{Hidaka:2020izy}
Y.~Hidaka, M.~Nitta, and R.~Yokokura, ``{Global 3-group symmetry and 't Hooft
  anomalies in axion electrodynamics},''
  \href{http://arxiv.org/abs/2009.14368}{{\ttfamily arXiv:2009.14368
  [hep-th]}}.

\bibitem{Yu:2020twi}
M.~Yu, ``{Symmetries and Anomalies of (1+1)d Theories: 2-groups and Symmetry
  Fractionalization},'' \href{http://arxiv.org/abs/2010.01136}{{\ttfamily
  arXiv:2010.01136 [hep-th]}}.

\bibitem{Maldacena:1997re}
J.~M. Maldacena, ``{The Large N limit of superconformal field theories and
  supergravity},'' \href{http://dx.doi.org/10.1023/A:1026654312961}{{\em Int.
  J. Theor. Phys.} {\bfseries 38} (1999) 1113--1133},
  \href{http://arxiv.org/abs/hep-th/9711200}{{\ttfamily arXiv:hep-th/9711200}}.

\bibitem{Gubser:1998bc}
S.~Gubser, I.~R. Klebanov, and A.~M. Polyakov, ``{Gauge theory correlators from
  noncritical string theory},''
  \href{http://dx.doi.org/10.1016/S0370-2693(98)00377-3}{{\em Phys. Lett. B}
  {\bfseries 428} (1998) 105--114},
  \href{http://arxiv.org/abs/hep-th/9802109}{{\ttfamily arXiv:hep-th/9802109}}.

\bibitem{Witten:1998qj}
E.~Witten, ``{Anti-de Sitter space and holography},''
  \href{http://dx.doi.org/10.4310/ATMP.1998.v2.n2.a2}{{\em Adv. Theor. Math.
  Phys.} {\bfseries 2} (1998) 253--291},
  \href{http://arxiv.org/abs/hep-th/9802150}{{\ttfamily arXiv:hep-th/9802150}}.

\bibitem{Grozdanov:2017kyl}
S.~Grozdanov and N.~Poovuttikul, ``{Generalised global symmetries in
  holography: magnetohydrodynamic waves in a strongly interacting plasma},''
  \href{http://dx.doi.org/10.1007/JHEP04(2019)141}{{\em JHEP} {\bfseries 04}
  (2019) 141}, \href{http://arxiv.org/abs/1707.04182}{{\ttfamily
  arXiv:1707.04182 [hep-th]}}.

\bibitem{Hofman:2017vwr}
D.~M. Hofman and N.~Iqbal, ``{Generalized global symmetries and holography},''
  \href{http://dx.doi.org/10.21468/SciPostPhys.4.1.005}{{\em SciPost Phys.}
  {\bfseries 4} no.~1, (2018) 005},
  \href{http://arxiv.org/abs/1707.08577}{{\ttfamily arXiv:1707.08577
  [hep-th]}}.

\bibitem{Grozdanov:2018ewh}
S.~Grozdanov and N.~Poovuttikul, ``{Generalized global symmetries in states
  with dynamical defects: The case of the transverse sound in field theory and
  holography},'' \href{http://dx.doi.org/10.1103/PhysRevD.97.106005}{{\em Phys.
  Rev. D} {\bfseries 97} no.~10, (2018) 106005},
  \href{http://arxiv.org/abs/1801.03199}{{\ttfamily arXiv:1801.03199
  [hep-th]}}.

\bibitem{Harlow:2018tng}
D.~Harlow and H.~Ooguri, ``{Symmetries in quantum field theory and quantum
  gravity},'' \href{http://arxiv.org/abs/1810.05338}{{\ttfamily
  arXiv:1810.05338 [hep-th]}}.

\bibitem{Bergman:2020ifi}
O.~Bergman, Y.~Tachikawa, and G.~Zafrir, ``{Generalized symmetries and
  holography in ABJM-type theories},''
  \href{http://dx.doi.org/10.1007/JHEP07(2020)077}{{\em JHEP} {\bfseries 07}
  (2020) 077}, \href{http://arxiv.org/abs/2004.05350}{{\ttfamily
  arXiv:2004.05350 [hep-th]}}.

\bibitem{Witten:2003ya}
E.~Witten, ``{SL(2,Z) action on three-dimensional conformal field theories with
  Abelian symmetry},'' \href{http://arxiv.org/abs/hep-th/0307041}{{\ttfamily
  arXiv:hep-th/0307041}}.

\bibitem{Iqbal:2020lrt}
N.~Iqbal and N.~Poovuttikul, ``{2-group global symmetries, hydrodynamics and
  holography},'' \href{http://arxiv.org/abs/2010.00320}{{\ttfamily
  arXiv:2010.00320 [hep-th]}}.

\bibitem{Bianchi:2001de}
M.~Bianchi, D.~Z. Freedman, and K.~Skenderis, ``{How to go with an RG flow},''
  \href{http://dx.doi.org/10.1088/1126-6708/2001/08/041}{{\em JHEP} {\bfseries
  08} (2001) 041}, \href{http://arxiv.org/abs/hep-th/0105276}{{\ttfamily
  arXiv:hep-th/0105276}}.

\bibitem{Bianchi:2001kw}
M.~Bianchi, D.~Z. Freedman, and K.~Skenderis, ``{Holographic
  renormalization},''
  \href{http://dx.doi.org/10.1016/S0550-3213(02)00179-7}{{\em Nucl. Phys. B}
  {\bfseries 631} (2002) 159--194},
  \href{http://arxiv.org/abs/hep-th/0112119}{{\ttfamily arXiv:hep-th/0112119}}.

\bibitem{deHaro:2000vlm}
S.~de~Haro, S.~N. Solodukhin, and K.~Skenderis, ``{Holographic reconstruction
  of space-time and renormalization in the AdS / CFT correspondence},''
  \href{http://dx.doi.org/10.1007/s002200100381}{{\em Commun. Math. Phys.}
  {\bfseries 217} (2001) 595--622},
  \href{http://arxiv.org/abs/hep-th/0002230}{{\ttfamily arXiv:hep-th/0002230}}.

\bibitem{Skenderis:2002wp}
K.~Skenderis, ``{Lecture notes on holographic renormalization},''
  \href{http://dx.doi.org/10.1088/0264-9381/19/22/306}{{\em Class. Quant.
  Grav.} {\bfseries 19} (2002) 5849--5876},
  \href{http://arxiv.org/abs/hep-th/0209067}{{\ttfamily arXiv:hep-th/0209067}}.

\bibitem{DeWolfe:2018dkl}
O.~DeWolfe, ``{TASI Lectures on Applications of Gauge/Gravity Duality},''
  \href{http://dx.doi.org/10.22323/1.305.0014}{{\em PoS} {\bfseries TASI2017}
  (2018) 014}, \href{http://arxiv.org/abs/1802.08267}{{\ttfamily
  arXiv:1802.08267 [hep-th]}}.

\bibitem{Papadimitriou:2004ap}
I.~Papadimitriou and K.~Skenderis, ``{AdS / CFT correspondence and geometry},''
  \href{http://dx.doi.org/10.4171/013-1/4}{{\em IRMA Lect. Math. Theor. Phys.}
  {\bfseries 8} (2005) 73--101},
  \href{http://arxiv.org/abs/hep-th/0404176}{{\ttfamily arXiv:hep-th/0404176}}.

\bibitem{Papadimitriou:2010as}
I.~Papadimitriou, ``{Holographic renormalization as a canonical
  transformation},'' \href{http://dx.doi.org/10.1007/JHEP11(2010)014}{{\em
  JHEP} {\bfseries 11} (2010) 014},
  \href{http://arxiv.org/abs/1007.4592}{{\ttfamily arXiv:1007.4592 [hep-th]}}.

\bibitem{Klebanov:1999tb}
I.~R. Klebanov and E.~Witten, ``{AdS / CFT correspondence and symmetry
  breaking},'' \href{http://dx.doi.org/10.1016/S0550-3213(99)00387-9}{{\em
  Nucl. Phys. B} {\bfseries 556} (1999) 89--114},
  \href{http://arxiv.org/abs/hep-th/9905104}{{\ttfamily arXiv:hep-th/9905104}}.

\bibitem{Aharony:2001pa}
O.~Aharony, M.~Berkooz, and E.~Silverstein, ``{Multiple trace operators and
  nonlocal string theories},''
  \href{http://dx.doi.org/10.1088/1126-6708/2001/08/006}{{\em JHEP} {\bfseries
  08} (2001) 006}, \href{http://arxiv.org/abs/hep-th/0105309}{{\ttfamily
  arXiv:hep-th/0105309}}.

\bibitem{Witten:2001ua}
E.~Witten, ``{Multitrace operators, boundary conditions, and AdS / CFT
  correspondence},'' \href{http://arxiv.org/abs/hep-th/0112258}{{\ttfamily
  arXiv:hep-th/0112258}}.

\bibitem{Berkooz:2002ug}
M.~Berkooz, A.~Sever, and A.~Shomer, ``{'Double trace' deformations, boundary
  conditions and space-time singularities},''
  \href{http://dx.doi.org/10.1088/1126-6708/2002/05/034}{{\em JHEP} {\bfseries
  05} (2002) 034}, \href{http://arxiv.org/abs/hep-th/0112264}{{\ttfamily
  arXiv:hep-th/0112264}}.

\bibitem{Papadimitriou:2007sj}
I.~Papadimitriou, ``{Multi-Trace Deformations in AdS/CFT: Exploring the Vacuum
  Structure of the Deformed CFT},''
  \href{http://dx.doi.org/10.1088/1126-6708/2007/05/075}{{\em JHEP} {\bfseries
  05} (2007) 075}, \href{http://arxiv.org/abs/hep-th/0703152}{{\ttfamily
  arXiv:hep-th/0703152}}.

\bibitem{Leigh:2003gk}
R.~G. Leigh and A.~C. Petkou, ``{Holography of the N=1 higher spin theory on
  AdS(4)},'' \href{http://dx.doi.org/10.1088/1126-6708/2003/06/011}{{\em JHEP}
  {\bfseries 06} (2003) 011},
  \href{http://arxiv.org/abs/hep-th/0304217}{{\ttfamily arXiv:hep-th/0304217}}.

\bibitem{Leigh:2003ez}
R.~G. Leigh and A.~C. Petkou, ``{SL(2,Z) action on three-dimensional CFTs and
  holography},'' \href{http://dx.doi.org/10.1088/1126-6708/2003/12/020}{{\em
  JHEP} {\bfseries 12} (2003) 020},
  \href{http://arxiv.org/abs/hep-th/0309177}{{\ttfamily arXiv:hep-th/0309177}}.

\bibitem{Petkou:2004nu}
A.~C. Petkou, ``{Holography, duality and higher-spin theories},'' in {\em {1st
  Solvay Workshop on Higher Spin Gauge Theories}}, pp.~65--78.
\newblock 10, 2004.
\newblock \href{http://arxiv.org/abs/hep-th/0410116}{{\ttfamily
  arXiv:hep-th/0410116}}.

\bibitem{Yee:2004ju}
H.-U. Yee, ``{A Note on AdS / CFT dual of SL(2,Z) action on 3-D conformal field
  theories with U(1) symmetry},''
  \href{http://dx.doi.org/10.1016/j.physletb.2004.05.082}{{\em Phys. Lett. B}
  {\bfseries 598} (2004) 139--148},
  \href{http://arxiv.org/abs/hep-th/0402115}{{\ttfamily arXiv:hep-th/0402115}}.

\bibitem{Marolf:2006nd}
D.~Marolf and S.~F. Ross, ``{Boundary Conditions and New Dualities: Vector
  Fields in AdS/CFT},''
  \href{http://dx.doi.org/10.1088/1126-6708/2006/11/085}{{\em JHEP} {\bfseries
  11} (2006) 085}, \href{http://arxiv.org/abs/hep-th/0606113}{{\ttfamily
  arXiv:hep-th/0606113}}.

\bibitem{Witten:1998wy}
E.~Witten, ``{AdS / CFT correspondence and topological field theory},''
  \href{http://dx.doi.org/10.1088/1126-6708/1998/12/012}{{\em JHEP} {\bfseries
  12} (1998) 012}, \href{http://arxiv.org/abs/hep-th/9812012}{{\ttfamily
  arXiv:hep-th/9812012}}.

\bibitem{Belov:2004ht}
D.~Belov and G.~W. Moore, ``{Conformal blocks for AdS(5) singletons},''
  \href{http://arxiv.org/abs/hep-th/0412167}{{\ttfamily arXiv:hep-th/0412167}}.

\end{thebibliography}\endgroup

\end{document}